\documentclass{rnaastex}
\usepackage{graphicx}

\begin{document}

\title{Another unWISE Update: The Deepest Ever Full-sky Maps at 3$-$5 Microns}

\correspondingauthor{Aaron M. Meisner}

\author{A.~M. Meisner}
\affiliation{Berkeley Center for Cosmological Physics, Berkeley, CA 94720, USA}
\affiliation{Lawrence Berkeley National Laboratory, Berkeley, CA, 94720, USA}

\author{D. Lang}
\affiliation{Department of Astronomy \& Astrophysics and Dunlap Institute, University of Toronto, Toronto, ON M5S 3H4, Canada}
\affiliation{Department of Physics \& Astronomy, University of Waterloo, 200 University Avenue West, Waterloo, ON, N2L 3G1, Canada}

\author{D.~J. Schlegel}
\affiliation{Lawrence Berkeley National Laboratory, Berkeley, CA, 94720, USA}

\keywords{atlases --- infrared: general --- methods: data analysis --- surveys  --- techniques: image processing}

\section{} 

With its large field of view and high sensitivity at mid-infrared wavelengths, the Wide-field Infrared Survey Explorer \citep[WISE;][]{wright10} is a unique and 
powerful tool for numerous applications in Galactic and extragalactic astrophysics, from nearby brown dwarfs to precision cosmology. WISE acquired 3.4$\mu$m (W1) and 4.6$\mu$m (W2) data throughout 2010 before entering a $\sim$3 year hibernation period, then recommenced surveying as the NEOWISE-Reactivation \citep[NEOWISER;][]{neowiser} asteroid-hunting mission in late 2013. Observations in W1 and W2 have continued ever since. Although NEOWISER has supplied over 75\% of archival W1 and W2 observations, the mission itself does not provide any coadded data products optimized for science beyond the inner solar system.

We have undertaken an effort to repurpose NEOWISER exposures for Galactic and extragalactic studies, by creating full-sky coadds combining all W1 and W2 exposures ever acquired \citep{fulldepth_neo2, meisner16, tr_neo2}. Our custom W1/W2 reprocessings generate two types of output maps: (1) ``full-depth'' coadds which stack all data together, maximizing static source signal-to-noise, and (2) a set of ``time-resolved'' coadds spaced at six-month intervals, enabling motion and variability measurements of faint WISE sources. In both cases we preserve the native WISE angular resolution, performing coaddition with an adaptation of the unWISE pipeline \citep{lang14}. 

Here we update our full-depth coadds by folding in the most recently published year of W1/W2 exposures released by NEOWISER. These new single-frame data were acquired between 2015 December 13 and 2016 December 13, and became public in 2017 June. In the present work, we simply re-ran the latest unWISE coaddition code \citep{fulldepth_neo2} on inputs including this additional year of publicly available NEOWISER frames. The resulting set of full-depth coadds uniformly incorporates all publicly available W1 and W2 exposures, with observation dates ranging from 2010 January 7 to 2016 December 13. The inputs consisted of $\sim$10.5 million frames per band, totaling $\sim$140 terabytes of single-exposure pixel data. 

The mean integer coverage is 143 (142) frames per sky location in W1 (W2), roughly four times that of the AllWISE Atlas stacks. The minimum integer coverage is 49 (43) frames in W1 (W2) -- there are no missing ``holes.'' The maximum integer coverage is 21,381 (21,340) frames in W1 (W2), occurring near the north ecliptic pole. Our computations did not require us to discard any usable data at the ecliptic poles.

\begin{figure}[h!]
\begin{center}
\includegraphics[width=7.0in]{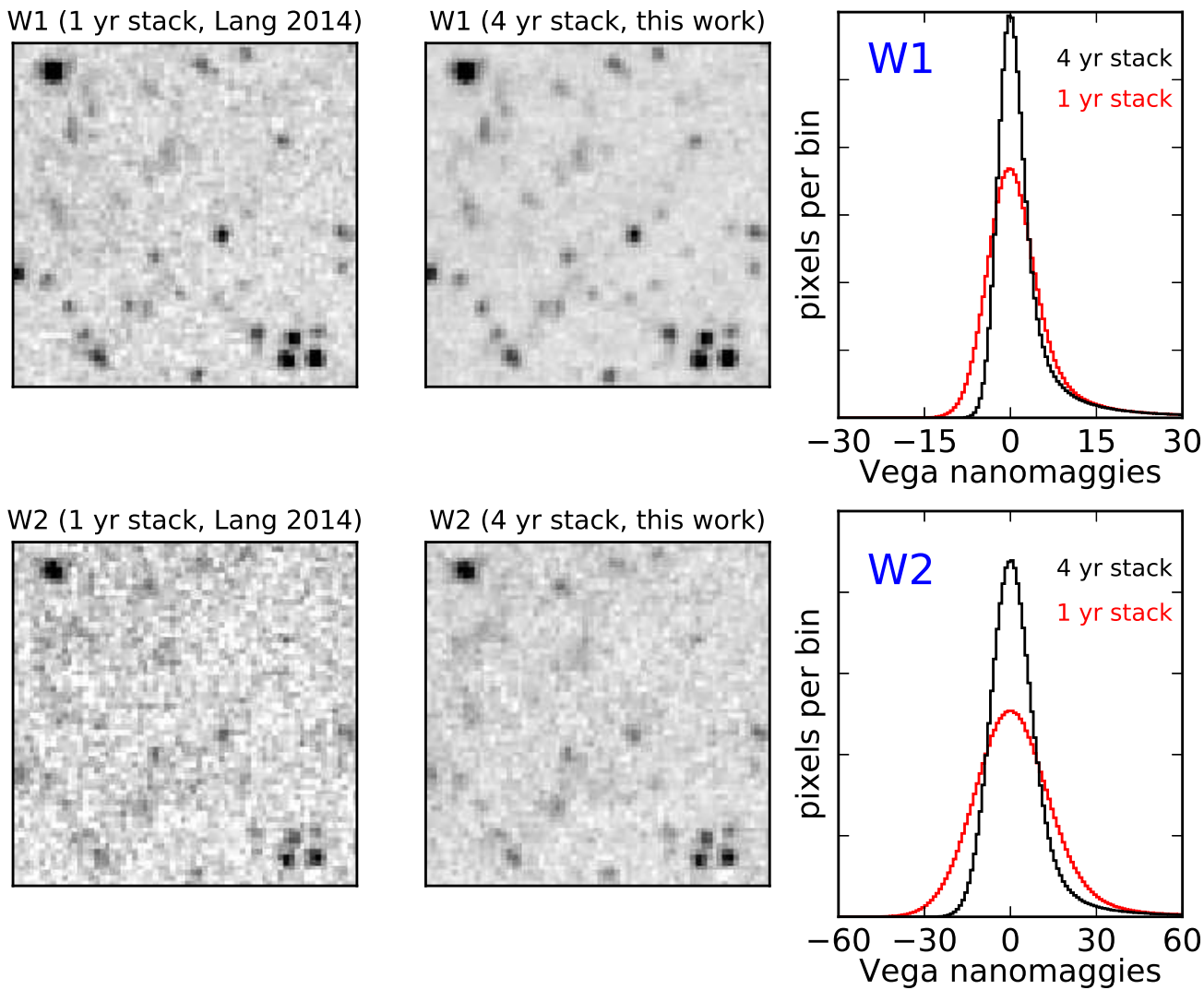}
\caption{Illustration of the improved depth of the present W1/W2 coadds relative to those which only incorporate pre-hibernation WISE imaging. The top (bottom) row shows W1 (W2). Left column: \cite{lang14} coadds. Center column: Corresponding coadds from this work. Cutouts are 2.9$'$$\times$2.9$'$, centered at $(\alpha, \delta)$ = (149.6155$^{\circ}$, 1.4510$^{\circ}$). Right column: Pixel value histograms for a $\sim$2.5 deg$^2$ region from which these cutouts have been drawn, indicating that a $\sim$1.5-2$\times$ reduction in statistical noise has been achieved.\label{fig:1}}
\end{center}
\end{figure}

Our new coadds\footnote{Available at \url{http://unwise.me/fulldepth_neo3}.} represent the deepest ever full-sky maps at 3$-$5$\mu$m, and constitute a vital input for selecting the Dark Energy Spectroscopic Instrument's luminous red galaxy and quasar targets \citep{desi, desi_part2, desi_part1}. Figure \ref{fig:1} illustrates the $\sim$1.5-2$\times$ reduced pixel noise achieved by our latest coadds, which incorporate $\sim$4 years of W1/W2 observations, relative to coadds including only pre-hibernation data, such as the \cite{lang14} unWISE coadds and AllWISE Atlas stacks. We will continue updating our full-depth and time-resolved unWISE coadds as more NEOWISER data become available.

\acknowledgments

This work has been supported by NASA ADAP grant NNH17AE75I.

This research makes use of data products from WISE, which is a joint project of UCLA, and JPL/Caltech, funded by NASA. This research also makes use of data products from NEOWISE, which is a project of JPL/Caltech, funded by the Planetary Science Division of NASA. This research has made use of the NASA/IPAC Infrared Science Archive, which is operated by JPL/Caltech, under contract with NASA.

The National Energy Research Scientific Computing Center, which is supported by the Office of Science of the U.S. Department of Energy under Contract No. DE-AC02-05CH11231, provided staff, computational resources, and data storage for this project.

\bibliography{fulldepth_neo3}

\end{document}